\documentclass[aps,prb,superscriptaddress,amsmath,amssymb,twocolumn]{revtex4-1}

\usepackage{amsmath}
\usepackage{graphicx}
\usepackage{amssymb}
\usepackage[usenames,dvipsnames]{xcolor}
\usepackage{subfigure}
\usepackage{dcolumn}
\usepackage{bm}
\usepackage{ulem}
\makeatletter
\begin{document}
\newcommand{\mr}[1]{{\color{red}#1}}

\title{How to remove the spurious resonances from ring polymer molecular dynamics}
\author{Mariana Rossi}
\affiliation {Physical and Theoretical Chemistry Laboratory, 
University of Oxford, South Parks Road, Oxford OX1 3QZ, United Kingdom}
\author{Michele Ceriotti}
\affiliation {Laboratory of Computational Science and Modeling, IMX, \'Ecole Polytechnique F\'ed\'erale de Lausanne, 1015 Lausanne, Switzerland}
\author{David E. Manolopoulos}
\affiliation {Physical and Theoretical Chemistry Laboratory, 
University of Oxford, South Parks Road, Oxford OX1 3QZ, United Kingdom}

\begin{abstract} 
Two of the most successful methods that are presently available for simulating the quantum dynamics of condensed phase systems are centroid molecular dynamics (CMD) and ring polymer molecular dynamics (RPMD). Despite their conceptual differences, practical implementations of these methods differ in just two respects: the choice of the Parrinello-Rahman mass matrix and whether or not a thermostat is applied to the internal modes of the ring polymer during the dynamics. Here we explore a method which is halfway between the two approximations: we keep the path integral bead masses equal to the physical particle masses but attach a Langevin thermostat to the internal modes of the ring polymer during the dynamics. We justify this by showing analytically that the inclusion of an internal mode thermostat does not affect any of the desirable features of RPMD:  thermostatted RPMD (TRPMD) is equally valid with respect to everything that has actually been proven about the method as RPMD itself. In particular, because of the choice of bead masses, the resulting method is still optimum in the short-time limit, and the transition state approximation to its reaction rate theory remains closely related to the semiclassical instanton approximation in the deep quantum tunneling regime. In effect, there is a continuous family of methods with these properties, parameterised by the strength of the Langevin friction. Here we explore numerically how the approximation to quantum dynamics depends on this friction, with a particular emphasis on vibrational spectroscopy. We find that a broad range of frictions approaching optimal damping give similar results, and that these results are immune to both the resonance problem of RPMD and the curvature problem of CMD.
\end{abstract}

\maketitle

\section{Introduction}

Centroid molecular dynamics\cite{cao94a,jang99} (CMD) and ring polymer molecular dynamics\cite{craig04,braams06b} (RPMD) are two closely related approximate techniques for including quantum mechanical zero point energy and tunnelling effects in molecular dynamics simulations of condensed phase systems. Both are based on the imaginary time path integral formulation of quantum statistical mechanics,\cite{feynman65} but in different ways: RPMD is classical molecular dynamics in the extended phase space of the imaginary time path integral (or ring polymer\cite{chandler81}), whereas CMD is classical molecular dynamics on the potential of mean force generated by the thermal fluctuations of this ring polymer around its centroid.

Both of these techniques are now well established and have been used in a wide variety of applications.\cite{voth96,habershon13} They are especially useful for calculating diffusion coefficients and other transport properties of liquids,\cite{pavese96,lobaugh97,yonetani03,hone04,miller05a,miller05b,hone06,markland08,miller08,habershon09,menzeleev10,markland11} and for including zero-point energy and tunnelling effects in the calculation of chemical reaction rates.\cite{voth93,cao96,cao97,jang00,geva01,craig05a,craig05b,collepardo08,collepardo09,suleimanov11,menzeleev11,boekelheide11,tudela12,allen13,li14,menzeleev14} In these contexts, both methods appear to be reasonably reliable, on the basis of comparisons with full quantum mechanical results in cases where these are available,\cite{geva01,craig05a,craig05b,collepardo09,suleimanov11,tudela12} and exact quantum mechanical moment constraints in cases where they are not.\cite{miller05a,miller05b,braams06a,habershon07,perez09} One explanation for why RPMD works well for these problems is that diffusion coefficients and reaction rate coefficients are determined by the zero-frequency components of the relevant (velocity and reactive flux) autocorrelation spectra, and are therefore comparatively insensitive to the artificial high-frequency dynamics of the ring polymer internal modes.\cite{habershon13}

There are however some important situations in which RPMD and CMD are known to give unphysical results, the classic example being in the simulation of vibrational spectroscopy. Here RPMD fails because of the so-called {\em resonance problem}:\cite{shiga08,habershon08,witt09} when the frequency of a physical vibration (such as an O-H stretch in liquid water) comes into resonance with an excited internal mode of the ring polymer in another degree of freedom (such as a librational mode of the liquid), the resulting interaction causes a spurious splitting of the physical peak in the calculated spectrum.\cite{habershon08} In a simulation at room temperature, the first internal excitation of the ring polymer occurs at a wavenumber of $kT/\hbar c \sim 1300$ cm$^{-1}$, and so any spectral feature beyond this wavenumber could in principle be affected by this problem.

The resonance problem does not occur in CMD because the high-frequency vibrations of the internal modes of the ring polymer are averaged over in this method to calculate the centroid potential of mean force: they do not appear as dynamical variables. However, CMD can also give unphysical vibrational spectra, for a different reason known as the {\em curvature problem}.\cite{witt09,ivanov10} This problem is again inherently multidimensional and is expected to arise whenever a high frequency stretching coordinate is combined with an angular coordinate, such as a rotational or a torsional degree of freedom. At low temperatures, the ring polymer spreads out around the angular coordinate, resulting in a softening of the centroid potential of mean force along the stretch.\cite{witt09,ivanov10}  Because of this, the stretching peak in the simulated spectrum is artificially red-shifted and broadened. In the extreme case of a combination with a free rotation, the frequency of the stretching peak can even tend to zero as $T\to 0$.\cite{witt09}

In this paper, we shall show that both of these problems can be avoided by adopting a method that is halfway between RPMD and CMD. Despite their conceptual differences, practical implementations of these two methods are very similar. In RPMD, the masses of the ring polymer beads are chosen to be the physical particle masses, and the dynamics that is used to calculate correlation functions is microcanonical.\cite{craig04} In the adiabatic implementation of CMD,\cite{cao94b} the internal modes of the ring polymer are given much smaller masses so that they vibrate at frequencies well beyond the physical spectral range, and a thermostat is attached to these internal modes to generate the canonical centroid potential of mean force. We shall show here that, in the context of vibrational spectroscopy, attaching a thermostat to the internal modes of the ring polymer during the dynamics is a good idea, because it damps out the spurious resonances from RPMD. However, adiabatically separating the internal vibrations of the ring polymer from the centroid motion is a not such a good idea, because it is the origin of the curvature problem in CMD. 

The reason why it is legitimate to attach a thermostat to the internal modes of the ring polymer during the dynamics is discussed in Sec.~II: provided the thermostat only acts on the internal modes and not on the centroid, it does not have any impact on any of the established properties of RPMD. Sec.~III goes on to present some numerical examples to illustrate how attaching a thermostat to the internal modes while keeping the bead masses equal to the physical masses cures both the resonance problem of RPMD and the curvature problem of CMD, for a variety of model and more realistic vibrational problems. It also happens to be the case that thermostatted RPMD (TRPMD) provides the simplest computational scheme, because it avoids both the non-ergodicity of the microcanonical dynamics in RPMD and the need to use a small integration time step in CMD. Finally, to emphasise that it is not restricted to the calculation of vibrational spectra, we show that TRPMD gives a diffusion coefficient and orientational relaxation times for liquid water that are very similar to those obtained from RPMD and CMD. Sec.~IV contains some concluding remarks.

\section{Some properties of thermostatted RPMD}

\subsection{Standard ring polymer molecular dynamics}

Although RPMD can also be used to approximate cross correlation functions, we shall confine our attention here to Kubo-transformed autocorrelation functions of the form\cite{kubo57}
$$
\tilde{c}_{AA}(t) = {1\over\beta Z}\int_0^{\beta} d\lambda\, {\rm tr}\left[e^{-(\beta-\lambda)\hat{H}}
\hat{A}\,e^{-\lambda\hat{H}}\hat{A}(t)\right], \eqno(1)
$$
where 
$$
Z = {\rm tr}\left[e^{-\beta\hat{H}}\right], \eqno(2)
$$
and 
$$
\hat{A}(t) = e^{+i\hat{H}t/\hbar}\hat{A}\,e^{-i\hat{H}t/\hbar}. \eqno(3)
$$
Correlation functions of this form arise in a wide variety of contexts, including the theory of diffusion (in which $\hat{A}$ is the operator for the velocity of a tagged particle in a liquid), the theory of reaction rates (in which $\hat{A}$ is the operator for the flux of particles through a transition state dividing surface), and, most relevantly to the present study, the theory of vibrational spectroscopy (in which $\hat{A}$ is the dipole moment operator of the entire system).

Since the multi-dimensional generalisation is straightforward,\cite{habershon13} and it does not add anything other than more indices, we shall present all of the following equations for a model one-dimensional system with a Hamiltonian of the form
$$
\hat{H}={\hat{p}^2\over 2m}+V(\hat{q}). \eqno(4)
$$
Furthermore, since most physically interesting correlation functions involving non-local operators can be obtained by differentiating correlation functions of local operators (for example, the velocity autocorrelation function of a tagged particle in a liquid is minus the second time derivative of its position autocorrelation function), we shall restrict our attention to the case where $\hat{A}$ is a local Hermitian operator $A(\hat{q})$. 

The $n$-bead RPMD approximation to the correlation function in Eq.~(1) is then simply\cite{craig04}
$$
\tilde{c}_{AA}(t) = {1\over N_n}\int d{\bf p}\int d{\bf q}\,
e^{-\beta_nH_n({\bf p},{\bf q})}{A}_n({\bf q}){A}_n({\bf q}_t), \eqno(5)
$$
where $N_n=(2\pi\hbar)^nZ_n$ with
$$
Z_n = {1\over (2\pi\hbar)^n}\int d{\bf p}\int d{\bf q}\,e^{-\beta_nH_n({\bf p},{\bf q})}.  \eqno(6)
$$
Here $H_n({\bf p},{\bf q})$ is the classical Hamiltonian of a harmonic ring polymer with a potential of $V(q)$ acting on each bead,
$$
H_n({\bf p},{\bf q}) = \sum_{j=1}^n \left[{p_j^2\over 2m}+{1\over 2}m\omega_n^2(q_j-q_{j-1})^2+V(q_j)\right],\eqno(7)
$$
with $\omega_n=1/(\beta_n\hbar)$, $\beta_n=\beta/n$, and $q_0\equiv q_n$. The coordinates ${\bf q}_t= {\bf q}_t({\bf p},{\bf q})$ in Eq.~(5) are obtained from the classical dynamics generated by this Hamiltonian, 
$$
\dot{\bf p} = -{\partial H_n({\bf p},{\bf q})\over\partial {\bf q}}, \eqno(8)
$$
$$
\dot{\bf q} = +{\partial H_n({\bf p},{\bf q})\over\partial {\bf p}}, \eqno(9)
$$
and the functions $A_n({\bf q})$ and $A_n({\bf q}_t)$ are averaged over the beads of the ring polymer necklace at times $0$ and $t$,
$$
A_n({\bf q}) = {1\over n}\sum_{j=1}^n A(q_j). \eqno(10)
$$

The approximation in Eq.~(5) is clearly {\em ad hoc}, and no one has yet been able to derive it from first principles. However, it is consistent with the path integral description of static equilibrium properties, it gives a correlation function with the same time-reversal and detailed balance symmetries as the quantum correlation function\cite{craig04} (which in the present context combine to make $\tilde{c}_{AA}(t)$ a real and even function of $t$), it is exact for problems with continuous spectra in the high-temperature limit where the ring polymer shrinks to a single classical bead, it is as accurate as possible in the short-time limit for a method based on the classical evolution of imaginary-time path integral variables,\cite{braams06b} and it is exact at all times for the position autocorrelation function of a simple harmonic oscillator.\cite{craig04} 

Furthermore, Richardson and Althorpe\cite{richardson09} have recently established a close connection between RPMD transition state theory (TST) and the semiclassical instanton approximation in the deep quantum tunnelling regime, and Hele and Althorpe\cite{hele13a,althorpe13,hele13b} have gone on to show that RPMD TST is in fact the only known quantum mechanical version of TST with positive definite statistics that can be derived from the $t\to 0_+$ limit of a reactive flux-side correlation function. These important findings, which legitimise the use of RPMD for calculating chemical reaction rates more so than in any other context, are both clearly associated with the accuracy of the RPMD approximation in the short-time limit.

The key observation that motivated the present study is that, as we shall show next, all of the above features of RPMD continue to hold when the Hamiltonian dynamics of the ring polymer necklace is augmented with a Langevin thermostat, provided the thermostat only acts on the internal modes of the ring polymer and not on its centroid. In effect, RPMD is just one of a family of dynamical approximations that share the same correct symmetries and limiting cases, the family being parameterised by the strength of the Langevin friction. So by exploring the whole family, one might hope to find a method that alleviates some of the known problems of RPMD (such as the resonance problem mentioned in the Introduction), without sacrificing any of its established features.

\subsection{The effect of an internal mode thermostat}
 
The Hamiltonian equations of motion in Eqs.~(8) and~(9) can be written out more explicitly as
$$
\dot{\bf p} = {\bf F}({\bf q}) = {\bf f}({\bf q})-{\bf K}{\bf q}, \eqno(11)
$$
$$
\dot{\bf q} = {1\over m}{\bf p}. \eqno(12)
$$
Here ${\bf K}$ is a real, symmetric, positive semi-definite $n\times n$ spring constant matrix with elements
$$
K_{jj'} = m\omega_n^2\bigl(2\delta_{jj'}-\delta_{jj'-1}-\delta_{jj'+1}\bigr), \eqno(13)
$$
all indices being understood to be translated by a multiple of $n$ to lie between 1 and $n$, and ${\bf f}({\bf q})$ is a force vector with elements $f_j({\bf q})=-V'(q_j)$.

An important feature of the spring constant matrix is that it does not have any impact on the centroid 
$$
\bar{p} = {\bf e}^{\rm T}{\bf p}, \qquad \bar{q} = {\bf e}^{\rm T}{\bf q} \eqno(14)
$$
of the ring polymer, but only on the ring polymer internal modes. To see this, it suffices to note that
$$
{\bf e}^{\rm T}{\bf K} = {\bf 0}^{\rm T} \hbox{   and   } {\bf K}\,{\bf e}={\bf 0}, \eqno(15)
$$
where ${\bf e}$ is a vector with elements $e_j=1/n$ and ${\bf 0}$ is a vector of zeroes. This feature was used repeatedly by Braams and Manolopoulos\cite{braams06b} to investigate the short-time limit of RPMD, and we shall use it again in this context below.

Using the same notation, suppose that we now attach a Langevin thermostat to the dynamics by replacing Eq.~(11) with
$$
\dot{\bf p} = {\bf F}({\bf q})-\bm{\gamma}{\bf p}+\sqrt{2m\bm{\gamma}\over\beta_n}\bm{\xi}(t),
\eqno(16)
$$
where $\bm{\gamma}$ is a real, symmetric, positive semi-definite $n\times n$ friction matrix and $\bm{\xi}(t)$ is a vector of uncorrelated normal deviates with unit variance and zero mean ($\left<\xi_j(t)\right>=0$ and $\left<\xi_j(0)\xi_{j'}(t)\right>=\delta_{jj'}\delta(t)$). Then the thermostat will also be detached from the centroid (i.e., only act directly on the ring polymer internal modes) if ${\bf e}$ is an eigenvector of $\bm{\gamma}$ (and therefore also of $\sqrt{\bm{\gamma}}$) with eigenvalue zero,
$$
{\bf e}^{\rm T}\bm{\gamma} = {\bf 0}^{\rm T} \hbox{   and   } \bm{\gamma}\,{\bf e}={\bf 0}. \eqno(17)
$$
So let us consider whether any of the established properties of RPMD are affected by replacing Eq.~(11) with Eq.~(16) subject to this constraint on $\bm{\gamma}$.

First of all (and perhaps most importantly), the consistency with imaginary time path integral expressions for static equilibrium properties is maintained. One can either calculate these properties by averaging over short RPMD trajectories with initial conditions sampled from the Boltzmann distribution $e^{-\beta_nH_n({\bf p},{\bf q})}/N_n$, or by attaching a thermostat to the dynamics as in Eq.~(16). In fact, for the calculation of static equilibrium properties, attaching a thermostat to the dynamics is widely recognised as the better way to proceed.

The symmetry properties of RPMD are also unaffected by replacing Eq.~(11) with Eq.~(16), at least in the case that we are considering here: $\tilde{c}_{AA}(t)$ is still a real and even function of $t$. This follows because the Langevin dynamics is a stationary process -- a process that conserves the equilibrium distribution and is independent of the origin of time -- despite the appearance of $t$ in the noise term on the right-hand side of Eq.~(16). (The noise is uncorrelated from one instant of time to the next, and it has the same form at all instants of time.) 

In more detail, the autocorrelation function that results from the Langevin dynamics is 
$$
\tilde{c}_{AA}(t) = {1\over N_n}\int d{\bf p}\int d{\bf q}\,
e^{-\beta_nH_n({\bf p},{\bf q})}{A}_n({\bf q})\bar{A}_n({\bf q}_t), \eqno(18)
$$
where the overbar denotes an average over the stochastic process that takes the integration variables ${\bf p}$ and ${\bf q}$ to ${\bf q}_t({\bf p},{\bf q})$ in time $t>0$ (Langevin dynamics is only valid for positive times). By stationarity, this can be written equivalently as either
$$
\tilde{c}_{AA}(t) = {1\over N_n}\int d{\bf p}\int d{\bf q}\,
e^{-\beta_nH_n({\bf p},{\bf q})}\bar{A}_n({\bf q}_{t_0})\bar{A}_n({\bf q}_{t_0+t}), \eqno(19)
$$
or
$$
\tilde{c}_{AA}(t) = {1\over N_n}\int d{\bf p}\int d{\bf q}\,
e^{-\beta_nH_n({\bf p},{\bf q})}\bar{A}_n({\bf q}_{t_0-t})\bar{A}_n({\bf q}_{t_0}), \eqno(20)
$$
for any time origin $t_0>t$, and comparing these two equations we see that $\tilde{c}_{AA}(t)=\tilde{c}_{AA}(-t)$. Since $A(\hat{q})$ is Hermitian, $\tilde{c}_{AA}(t)$ is also real, and therefore a real and even function of $t$.

So far, we have not made any use of the constraint in Eq.~(17), which detaches the thermostat from the centroid of the ring polymer. However, this constraint is clearly needed for the dynamics to give the correct result in the high-temperature limit, and also for the position autocorrelation function of a harmonic oscillator. 

In the high temperature limit, where a single bead suffices to converge the autocorrelation function, the constraint makes $\bm{\gamma}$ zero, so the thermostat has no effect. One therefore recovers standard RPMD in this limit, which gives the correct (classical) result for systems with continuous spectra. As for the harmonic limit, when Eq.~(17) is satisfied the only term in Eq.~(16) that couples the centroid to the internal modes of the ring polymer is the force vector ${\bf f}({\bf q})$: in general, ${\bf e}^{\rm T}{\bf f}({\bf q})$ is not just a function of the centroid coordinate $\bar{q}={\bf e}^{\rm T}{\bf q}$. However, when the potential $V(q)$ is harmonic, $V(q)={1\over 2}m\omega^2q^2$, one has ${\bf e}^{\rm T}{\bf f}({\bf q})=-m\omega^2\bar{q}$, and the coupling of the centroid to the internal modes disappears. With the constraint in Eq.~(17), thermostatted RPMD will therefore give the same position ($\bar{q}$) autocorrelation function for a harmonic oscillator as RPMD.

Finally, let us consider the effect of the thermostat on the short-time limit of the correlation function $\tilde{c}_{AA}(t)$. Braams and Manolopoulos\cite{braams06b} have given a detailed analysis of the short-time limit of the standard RPMD approximation in Eq.~(5), in which they argued that it has an error of $O(t^8)$ for the position autocorrelation function and of $O(t^4)$ for a more general autocorrelation function of a non-linear operator $A(\hat{q})$. However, this argument assumed the odd time derivatives of the RPMD autocorrelation function to be continuous at $t=0$, as they are in quantum mechanics. By considering the special case of a simple harmonic oscillator, for which the autocorrelation function can be worked out analytically,\cite{horikoshi05} Jang {\em et al.}\cite{jang14} have shown that the third derivative of the RPMD approximation to $\tilde{c}_{q^2q^2}(t)$ becomes discontinuous at $t=0$ in the limit as $n\to\infty$ (the correlation function contains a term proportional to $|t|^3$ in this limit\cite{jang14}). This implies that the error in the RPMD approximation to this non-linear autocorrelation function is $O(t^3)$ rather than $O(t^4)$.\cite{jang14} Moreover a simple argument shows that the error in the RPMD position autocorrelation function will therefore be $O(t^7)$ rather than $O(t^8)$. This follows because $\tilde{c}^{(4)}_{qq}(t)=\tilde{c}_{ff}(t)$, where $f(q)=-V'(q)$ is the force. In an anharmonic potential, $f(q)$ is a non-linear function of $q$ (for example it could contain a component proportional to $q^2$), so the RPMD approximation to $\tilde{c}_{ff}(t)$ has an error of $O(t^3)$, and the approximation to $\tilde{c}_{qq}(t)$ an error of $O(t^{4+3})$.

To see how thermostatting the internal modes of the ring polymer affects these results, we shall now compare the first few initial time derivatives of the Langevin-thermostatted correlation function in Eq.~(18) with those of the standard RPMD correlation function in Eq.~(5). The $k$-th (forward) initial time derivative of $\tilde{c}_{AA}(t)$ is given by Eq.~(18) as
$$
\tilde{c}^{(k)}_{AA}(0) = {1\over N_n}\int d{\bf p}\int d{\bf q}\,
e^{-\beta_nH_n({\bf p},{\bf q})}{A}_n({\bf q})\bar{A}^{(k)}_n({\bf q}), \eqno(21)
$$
where $\bar{A}^{(k)}_n({\bf q})$ is the $k$-th time derivative of the noise-averaged $\bar{A}_n({\bf q}_t)$ at $t=0$. This is given by\cite{zwanzig01}
$$
\bar{A}_n^{(k)}({\bf q}) = \left(D^{\dagger}\right)^k A_n({\bf q}), \eqno(22)
$$
where $D^{\dagger}$ is the adjoint of the Fokker-Planck operator associated with the Langevin dynamics in Eqs.~(12) and~(16),
$$
D^{\dagger} = {{\bf p}\over m}\cdot{\partial\over\partial {\bf q}}+{\bf F}({\bf q})\cdot{\partial\over\partial {\bf p}}
- {\bf p}\cdot\bm{\gamma}\cdot{\partial\over\partial {\bf p}}+{m\over\beta_n}{\partial\over\partial{\bf p}}\cdot\bm{\gamma}\cdot {\partial\over\partial {\bf p}}. \eqno(23)
$$
The time derivatives of the standard RPMD autocorrelation are recovered by setting $\bm{\gamma}={\bf 0}$ in this operator, which converts it into the adjoint of the Liouville operator associated with the Hamiltonian dynamics in Eqs.~(11) and~(12). The goal is therefore to find the smallest value of $k$ for which $\tilde{c}^{(k)}_{AA}(0)$ involves any reference to $\bm{\gamma}$, which suffices to establish that the thermostatted and standard RPMD autocorrelation functions differ by $O(t^k)$.

In the special case of the position autocorrelation function, for which $A(\hat{q})=\hat{q}$ and $A_n({\bf q})={\bf e}^{\rm T}{\bf q}$, a straightforward but lengthy calculation that combines Eqs.~(21) to~(23) with the simplifications in Eqs.~(15) and~(17) gives 
$$
\tilde{c}^{(0)}_{qq}(0) = \left<{\bf e}^{\rm T}{\bf q}\,{\bf q}^{\rm T}{\bf e}\right>, \eqno(24a)
$$ 
$$
\tilde{c}^{(1)}_{qq}(0) = 0, \eqno(24b)
$$
$$
\tilde{c}^{(2)}_{qq}(0) = -{1\over \beta_n m}\left<{\bf e}^{\rm T}{\bf e}\right> \equiv -{1\over\beta m}, \eqno(24c)
$$
$$
\tilde{c}^{(3)}_{qq}(0) = 0, \eqno(24d)
$$
$$
\tilde{c}^{(4)}_{qq}(0) = +{1\over \beta_n m^2}\left< {\bf e}^{\rm T}{\bf H}({\bf q}){\bf e}\right>, \eqno(24e)
$$
$$
\tilde{c}^{(5)}_{qq}(0) = 0, \eqno(24f)
$$
$$
\tilde{c}^{(6)}_{qq}(0) = -{1\over \beta_n m^3}\left<{\bf e}^{\rm T}{\bf H}({\bf q})^2{\bf e}\right>, \eqno(24g)
$$
$$
\tilde{c}^{(7)}_{qq}(0) = +{1\over \beta_n m^3}\left<{\bf e}^{\rm T}{\bf H}({\bf q})\,\bm{\gamma}\,{\bf H}({\bf q}){\bf e}\right>. \eqno(24h)
$$
Here the angular brackets denote an equilibrium average
$$
\left<F({\bf q})\right> = {1\over N_n}\int d{\bf p}\int d{\bf q}\,e^{-\beta_nH_n({\bf p},{\bf q})}\,F({\bf q}), \eqno(25)
$$
and ${\bf H}({\bf q})$ is a diagonal matrix with diagonal elements $H_{jj}({\bf q})=V''(q_j)$.

When the potential $V(q)$ is harmonic, ${\bf e}^{\rm T}{\bf H}({\bf q})\bm{\gamma}{\bf H}({\bf q}){\bf e}$ is proportional to ${\bf e}^{\rm T}\bm{\gamma}{\bf e}$, which vanishes by virtue of Eq.~(17). All subsequent occurrences of $\bm{\gamma}$ in the higher derivatives $\tilde{c}^{(k)}_{qq}(0)$ also vanish for the same reason, so the thermostat has no effect on the position autocorrelation function. But when the potential $V(q)$ is anharmonic, there is no such simplification, and it follows from Eqs.~(24) that the difference between the thermostatted and standard RPMD autocorrelation functions is $O(t^7)$. Since this is exactly the same order as the error in the RPMD position autocorrelation function (see above), the short-time accuracy of the approximation is unaffected by attaching a thermostat to the internal modes of the ring polymer during the dynamics.

By contrast, changing the Parrinello-Rahman\cite{parrinello84} mass matrix so as to adiabatically separate the internal modes from the centroid mode, as is done in the adiabatic implementation of CMD,\cite{cao94c} certainly does have an effect on the short-time accuracy. The leading error is then $O(t^6)$ even before a thermostat is attached to the internal modes,\cite{braams06b} and the leading error in the fully converged CMD position autocorrelation function in which the centroid moves classically on its potential of mean force is only $O(t^4)$.\cite{cao94b}

In the more general case of the autocorrelation function of a non-linear operator $A(\hat{q})$, RPMD is considerably less accurate in the short-time limit,\cite{braams06b,jang14} and the same is true of thermostatted RPMD. In this case one finds from Eqs.~(21) to~(23) that the first time derivative $\tilde{c}_{AA}^{(k)}(0)$ to involve $\bm{\gamma}$ is
$$
\tilde{c}_{AA}^{(3)}(0) = +{1\over \beta_n m}\left<{\bf e}^{\rm T}{\bf A}'({\bf q})\,\bm{\gamma}\,{\bf A}'({\bf q}){\bf e}\right>, \eqno(26)
$$
where ${\bf A}'({\bf q})$ is a diagonal matrix with diagonal elements $A'(q_j)$. If $A(\hat{q})$ were a linear operator, these diagonal elements would all be the same, and $\tilde{c}_{AA}^{(3)}(0)$ would be zero by virtue of Eq.~(17). But for a non-linear operator, the right-hand side of Eq.~(26) is non-zero, so TRPMD differs from RPMD by a term of $O(t^3)$. Again, since this is the same order as the error in the RPMD autocorrelation function,\cite{jang14} nothing is lost by attaching a thermostat to the internal modes during the dynamics. (CMD is not really applicable to autocorrelation functions of non-linear operators, but it is nevertheless often used for this purpose in the form of the \lq\lq CMD with classical operators" approximation.\cite{voth96} The leading error is then $O(1)$ -- again 3 orders less accurate than RPMD.)

A closely related question is whether attaching a Langevin thermostat to the internal modes of the ring polymer will affect any of the results of Richardson and Althorpe\cite{richardson09} or Hele and Althorpe\cite{hele13a,althorpe13,hele13b} concerning  RPMD TST. The RPMD TST rate is obtained from the $t\to 0_+$ limit of a flux-side correlation function, which is (minus) the first time derivative of a side-side correlation function. The question is therefore whether the thermostat will affect the short-time limit of this side-side correlation function to $O(t)$. This can be investigated using Eqs.~(21) to (23) by replacing $A_n({\bf q})$ in Eq.~(10) with $A_n({\bf q})=h[f({\bf q})]$, where $h(x)$ is the Heaviside step function and $f({\bf q})=0$ is a transition state dividing surface in the ring polymer configuration space.\cite{richardson09,hele13a} Since this gives $\bar{A}_n^{(1)}({\bf q}) = \delta[f({\bf q})]\partial f({\bf q})/\partial {\bf q}\cdot{\bf p}/m$, which does not involve the friction matrix $\bm{\gamma}$, it is clear that attaching a thermostat to the dynamics will not have any effect on RPMD TST.

\subsection{Specification of the friction matrix}

So far, our discussion of TRPMD has been fairly general: we have not specified the friction matrix $\bm{\gamma}$ in Eq.~(16) beyond saying that it is real, symmetric, and positive semi-definite and that it satisfies the constraint in Eq.~(17). One could go further by considering a generalised Langevin equation thermostat in the form of a Langevin thermostat in an extended momentum space,\cite{ceriotti10b} which has the potential to provide an even milder perturbation to the Hamiltonian ring polymer dynamics. But that would lengthen the discussion considerably, and we have not yet explored this option in any detail. So instead, to make things more concrete, we shall now simply describe the particular family of friction matrices we have used for illustrative purposes in the calculations presented in Sec.~III. 

These friction matrices are based on the path integral Langevin equation (PILE) thermostat introduced in Ref.~\cite{ceriotti10}. The basic idea is to transform to the normal mode representation by diagonalising the harmonic spring constant matrix {\bf K} in Eq.~(13), and then to construct a diagonal friction matrix $\tilde{\bm{\gamma}}_{\rm c}$ in this representation. This has the advantage that the normal mode friction matrix can be chosen to give critical damping (optimum canonical sampling of the harmonic oscillator Hamiltonian) for each excited (non-centroid) normal mode of the free ring polymer.\cite{ceriotti10} To make the connection with the notation we have used above, $\tilde{\bm{\gamma}}_{\rm c}$ can then be transformed back into the ring-polymer bead representation to give a friction matrix for use in Eq.~(16). 

The relevant equations are simply
$$
{\bf C}^{\rm T}{\bf K}{\bf C} = m\tilde{\bm{\omega}}^2, \eqno(27)
$$
$$
\tilde{\bm{\gamma}}_{\rm c} = 2\tilde{\bm{\omega}}, \eqno(28)
$$
$$
\bm{\gamma}_{\rm c} = {\bf C}\,\tilde{\bm{\gamma}}_{\rm c}{\bf C}^{\rm T}, \eqno(29)
$$
where ${\bf C}$ is an orthogonal transformation matrix\cite{ceriotti10} and $\tilde{\bm{\omega}}$ is a diagonal matrix of free ring polymer vibrational frequencies: $\tilde{\omega}_{kk'} = 2\omega_n\sin(k\pi/n)\delta_{kk'}$ for $k,k'=0,1,\ldots,n-1$. 

Note that, since $\tilde{\omega}_{00}=0$, the friction matrix $\bm{\gamma}_{\rm c}$ does not have any effect on the centroid $(k=0)$ mode of the ring polymer, consistent with the constraint on $\bm{\gamma}$ in Eq.~(17). (In Ref.~\cite{ceriotti10}, a separate Langevin thermostat was applied to the centroid to facilitate the canonical sampling of static equilibrium properties, but to do so here would invalidate most of the results we have established for the autocorrelation functions obtained from thermostatted RPMD.) Note also that, while critical damping of the free ring polymer internal modes is certainly a good idea for the calculation of static equilibrium properties,\cite{ceriotti10} there is no {\em a priori} reason to suppose that it will be the best thing to do in thermostatted RPMD. We have therefore explored a range of dampings by defining the family of friction matrices
$$
\bm{\gamma} = \lambda{\bm{\gamma}}_{\rm c}, \eqno(30)
$$  
where $\lambda>1$ gives over-damping of the free ring polymer vibrations, $\lambda=1$ gives critical damping, $\lambda<1$ gives under-damping, and $\lambda=0$ recovers standard RPMD. The upshot is  that $\lambda=1/2$ is close to optimum for all of the problems we have studied. This is perhaps not surprising, because it is the choice that gives the most efficient canonical sampling of the configuration space of the free ring polymer (i.e., of its potential energy rather than its Hamiltonian).\cite{ceriotti10} In retrospect, $\lambda=1/2$ might also be better than $\lambda=1$ for the calculation of static equilibrium properties. We shall use this \lq optimal' value of $\lambda$ throughout the following section unless we specify otherwise.

\begin{figure}[ht]
\includegraphics[width=0.5\textwidth]{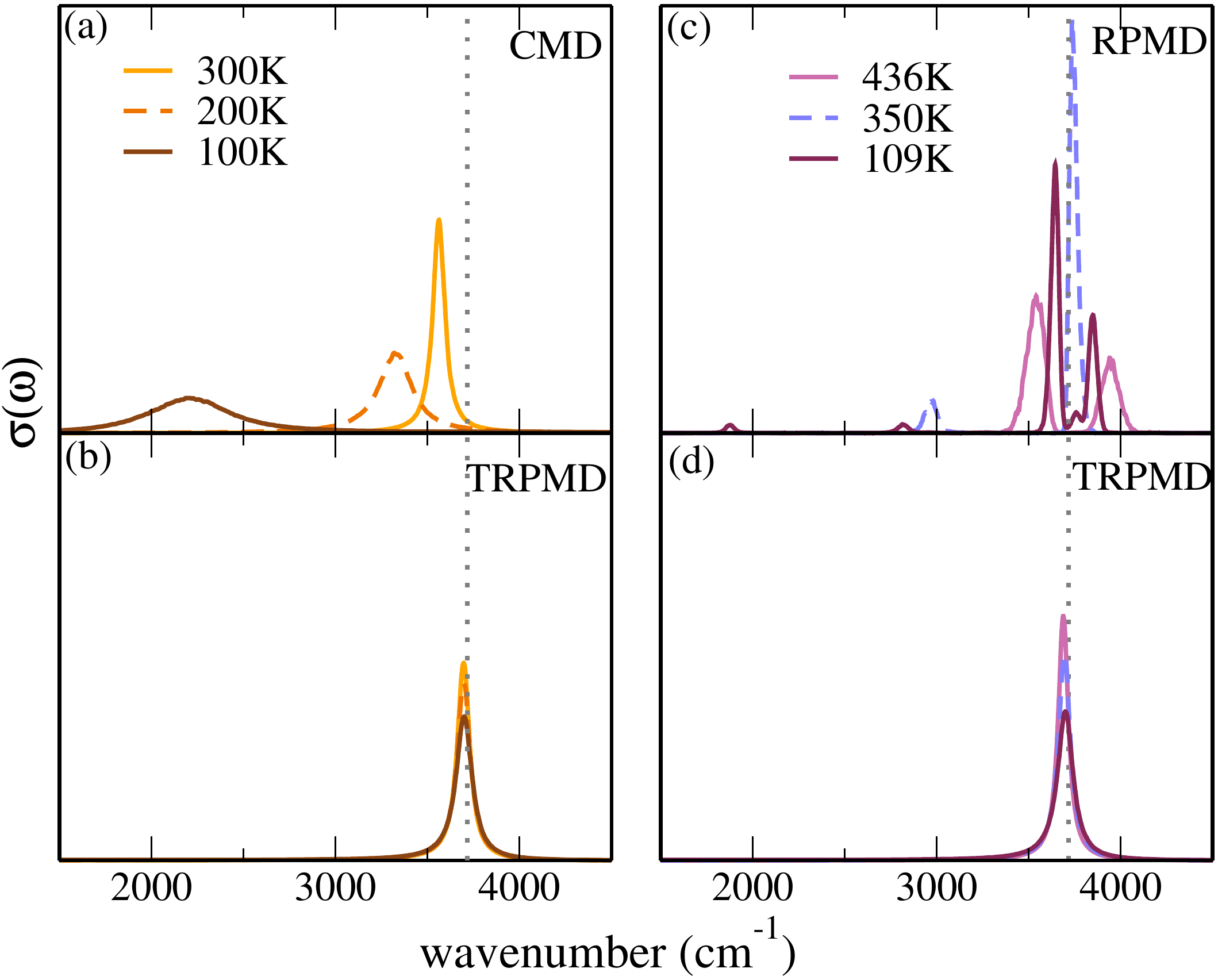}
\caption{Absorption cross sections for the harmonic OH molecule: (a) and (b) CMD and TRPMD methods at 100, 200, and 300 K; (c) and (d) RPMD and TRPMD methods at 109, 350, and 436 K. The dotted grey line indicates the correct harmonic vibrational frequency of the model, $\omega_\text{OH}=3715.6$ cm$^{-1}$.}
\end{figure}

\section{A solution to the resonance and curvature problems}

In the previous section we have shown that adding a stochastic thermostat to the internal degrees of freedom of the ring polymer without changing its mass matrix preserves all of the established properties of RPMD. Here we will demonstrate that, for several model and more realistic systems, the use of a PILE thermostat with optimal damping produces a method that is immune to both the resonance problem of RPMD and the curvature problem of CMD.

\subsection{Model molecules: harmonic OH and CH$_4$}

We shall start by considering two of the model molecules that were originally used to highlight the curvature and resonance problems.\cite{witt09} These are simplistic approximations to the OH and CH$_4$ molecules with harmonic interaction potentials
$$
V=\sum_{\rm bonds} \frac{k_b}{2} (r-r_e)^2 + \sum_{\rm angles} \frac{k_a}{2}(\theta - \theta_e)^2, \eqno(31)
$$
in which the values of $k_b$, $k_a$, $r_e$, and $\theta_e$ are given in Ref.~\cite{witt09}.

\begin{figure}[ht]
\includegraphics[width=0.5\textwidth]{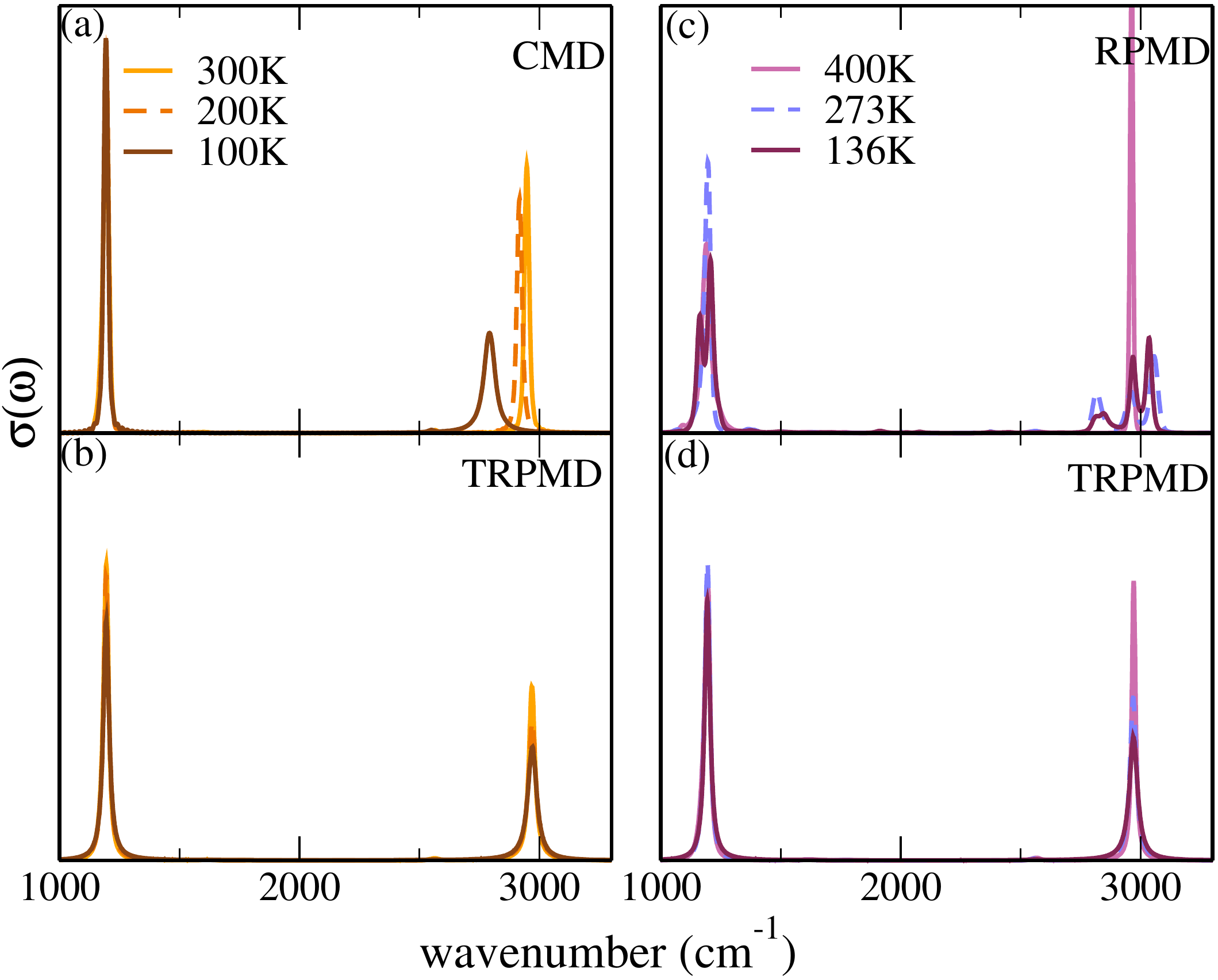}
\caption{Absorption cross sections for the harmonic CH$_4$ molecule: (a) and (b) CMD and TRPMD methods at 100, 200, and 300 K; (c) and (d) RPMD and TRPMD methods at 136, 273, and 400 K.}
\end{figure}

For each of these model molecules, we have calculated the dipole absorption cross section $\sigma(\omega)\propto \beta\omega^2\tilde{I}(\omega)$, where $\tilde{I}(\omega)$ is the Fourier transform of the Kubo-transformed autocorrelation function of the molecular dipole moment.~\cite{habershon08} RPMD, CMD, and TRPMD all yield approximations to this autocorrelation function. The resulting spectra are shown in Fig.~1 for the harmonic OH molecule and in Fig.~2 for CH$_4$. For the sake of comparison, these simulations were performed with translations and rotations removed using the procedure described in Ref.~\cite{witt09}, and at the same temperatures and with the same number of beads as in that study.

In both figures we show in panels (a) and (c) the CMD and RPMD spectra, which are fully consistent with the results reported in Ref.~\cite{witt09}. The curvature problem of CMD is apparent for both molecules. The high-frequency stretching peaks shift (quite dramatically) to lower frequencies and broaden significantly with decreasing temperature. The shift is more pronounced for the OH molecule than for the CH$_4$ molecule, because the presence of a harmonic bending term in the potential for methane reduces the spread of the ring polymer in the directions orthogonal to the stretch, thereby limiting the softening of the centroid potential of mean force along the stretching coordinate. For RPMD, one clearly sees that the peaks in the spectrum are split by resonances at temperatures where the free ring polymer frequencies coincide with the physical frequencies (109 K and 436 K for OH and 136 K and 273 K for CH$_4$). This resonance problem is just as severe for both molecules, because it is caused by the coupling between the physical vibrations of the stretching coordinate and comparatively low amplitude vibrations of the internal modes of the ring polymer in the directions orthogonal to the stretch.

\begin{figure}[htbp]
\centering
\includegraphics[width=0.45\textwidth]{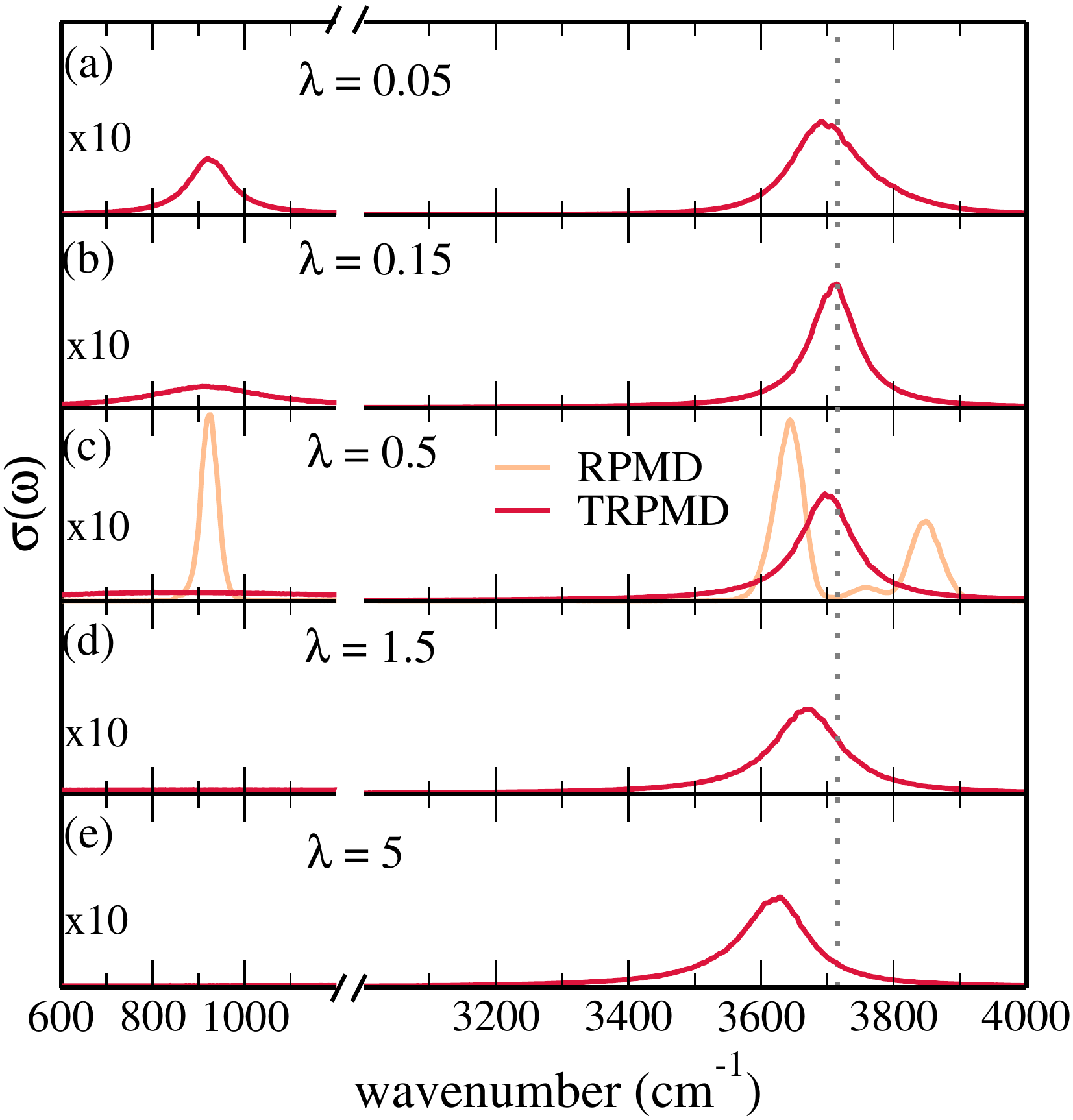}
\caption{Absorption cross sections for the harmonic OH 
molecule at 109 K. The various plots compare the RPMD spectrum to the 
TRPMD spectrum with different damping parameters $\lambda$. On the 
left side of the figure, the cross sections have been multiplied by a factor 10
for better visualisation.}
\end{figure}

In panels (b) and (d) we show the optimally-damped $(\lambda=1/2)$ TRPMD results for both molecules at the same temperatures used for the CMD and RPMD calculations. In all cases the resonant peak splitting disappears and the curvature problem is absent. This shows that the curvature problem originates from the choice of the Parrinello-Rahman mass matrix used in CMD, rather than from the thermostatting of the internal modes. As we will discuss more in detail in the next subsection, the peaks in the TRPMD spectrum are found to broaden as the temperature is lowered. This is because the internal vibrations of the ring polymer are damped by the Langevin friction but not eliminated. With optimal damping, they do not appear as sharp resonances, but they do lead to an indirect damping of the physical vibrations that becomes more pronounced as the temperature is lowered and the spacing between the internal frequencies of the ring polymer decreases.

\begin{figure}[ht]
\centering
\includegraphics[width=0.5\textwidth]{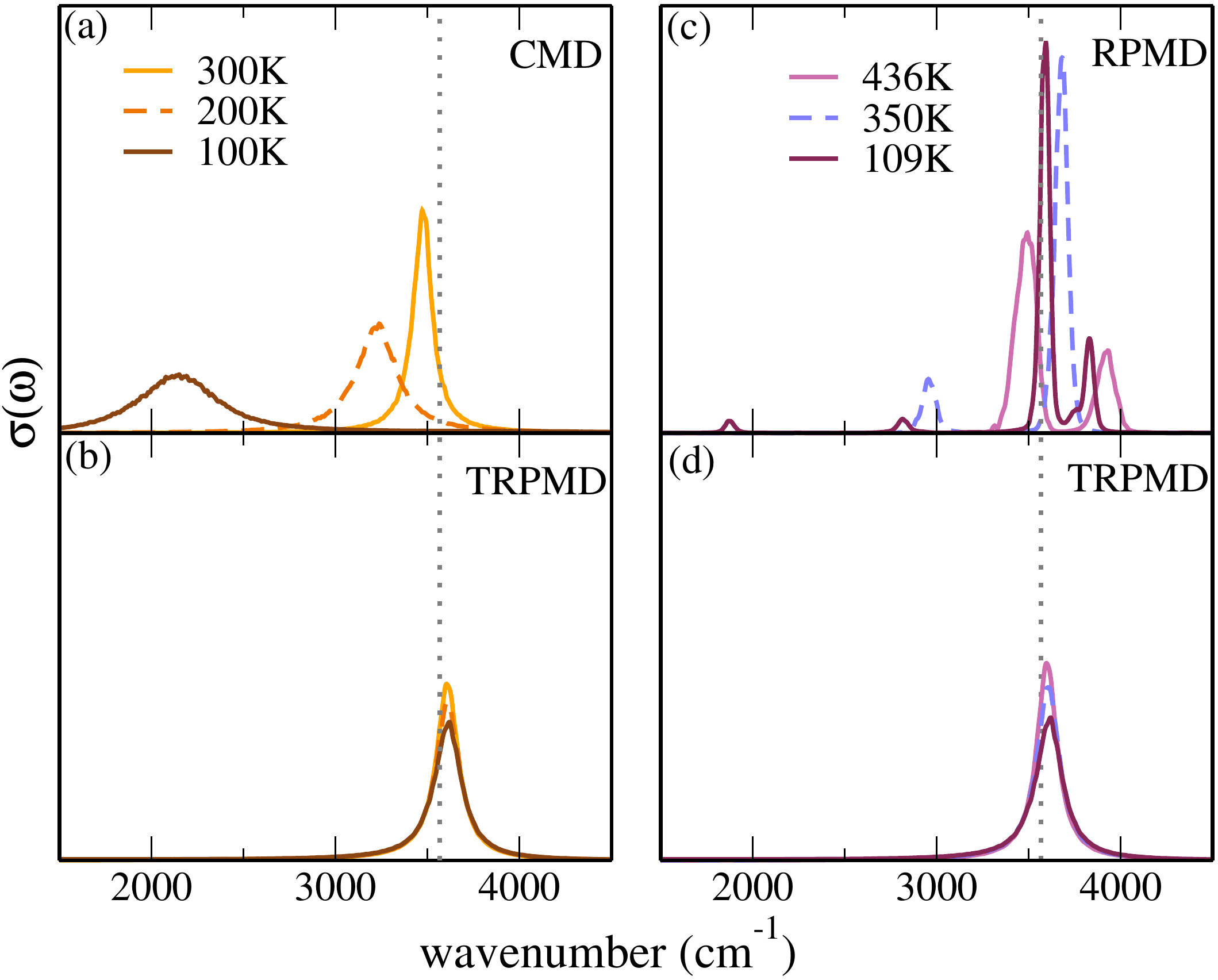}
\caption{Absorption cross sections for the anharmonic OH molecule: (a) and (b) CMD and TRPMD methods at 100, 200, and 300 K ; (c)  and (d) RPMD and TRPMD methods at 109, 350, and 436 K. The dotted grey line indicates the correct fundamental transition frequency $\omega_{1\leftarrow 0}=3568$ cm$^{-1}$.}
\label{fig:ohreal-rpmd-cmd-pile}
\end{figure}

To demonstrate that optimal damping is indeed the best choice for the TRPMD method, we have investigated the sensitivity of the results to the choice of the damping parameter. Figure~3 shows the TRPMD spectrum of the model OH molecule at 109 K calculated with different values of $\lambda$, in both the underdamped and overdamped regimes. The underdamped spectrum with $\lambda=0.15$ gives a peak very close to the correct OH stretching frequency, but it still shows a spurious peak at about 900 cm$^{-1}$ that corresponds to one of the internal excitations of the ring polymer in a direction orthogonal to the stretch. The strongly overdamped TRPMD spectrum with $\lambda=5$ shows no trace of this spurious peak, but the peak in the OH stretching region is now significantly red shifted (although not nearly by so much as it is at this temperature in CMD). Overall, optimal damping appears to be a reasonable compromise that washes out the spurious peak without dramatically red-shifting the physical vibration. Although the choice of $\lambda$ does affect the precise position of the stretching peak, there is a fairly large range of values approaching $\lambda=1/2$ for which the TRPMD spectrum does a reasonable job of capturing the correct peak position and masking the resonance artefact.

\subsection{Anharmonic OH}

As a second test case, we have considered an OH molecule with an anharmonic (Morse) interaction potential
$$
V = D_{e} \left[ 1-e^{-\alpha(r-r_{e})}\right]^2, \eqno(32)
$$ 
where $D_{e} = hc\,\omega_e^2/4\omega_ex_e$ and $\alpha=\sqrt{2\mu_{\rm OH}hc\,\omega_ex_e/\hbar^2}$ with $\omega_e=3737.76$ cm$^{-1}$, $\omega_ex_e=84.881$ cm$^{-1}$, and $r_{e} = 0.96966$ \AA.\cite{huber79} Since the vibrational energy levels of this potential are known exactly, this test case provides a convenient way to separate the curvature problem of CMD from the correct anharmonic red shift of the vibrational fundamental band, and also to study how the TRPMD method performs for a more realistic potential.

Figure~4 is analogous to Fig.~1, but now for the anharmonic OH molecule. For the CMD and RPMD methods (panels (a) and (c)) we see very similar curvature and resonance problems to those observed in the harmonic case. The position of the CMD stretching peak is noticeably red shifted with respect to the correct anharmonic result even at 300 K, and the RPMD stretching peak is plagued by resonances at all three temperatures considered in the figure.

\begin{figure}[ht]
\centering
\includegraphics[width=0.4\textwidth]{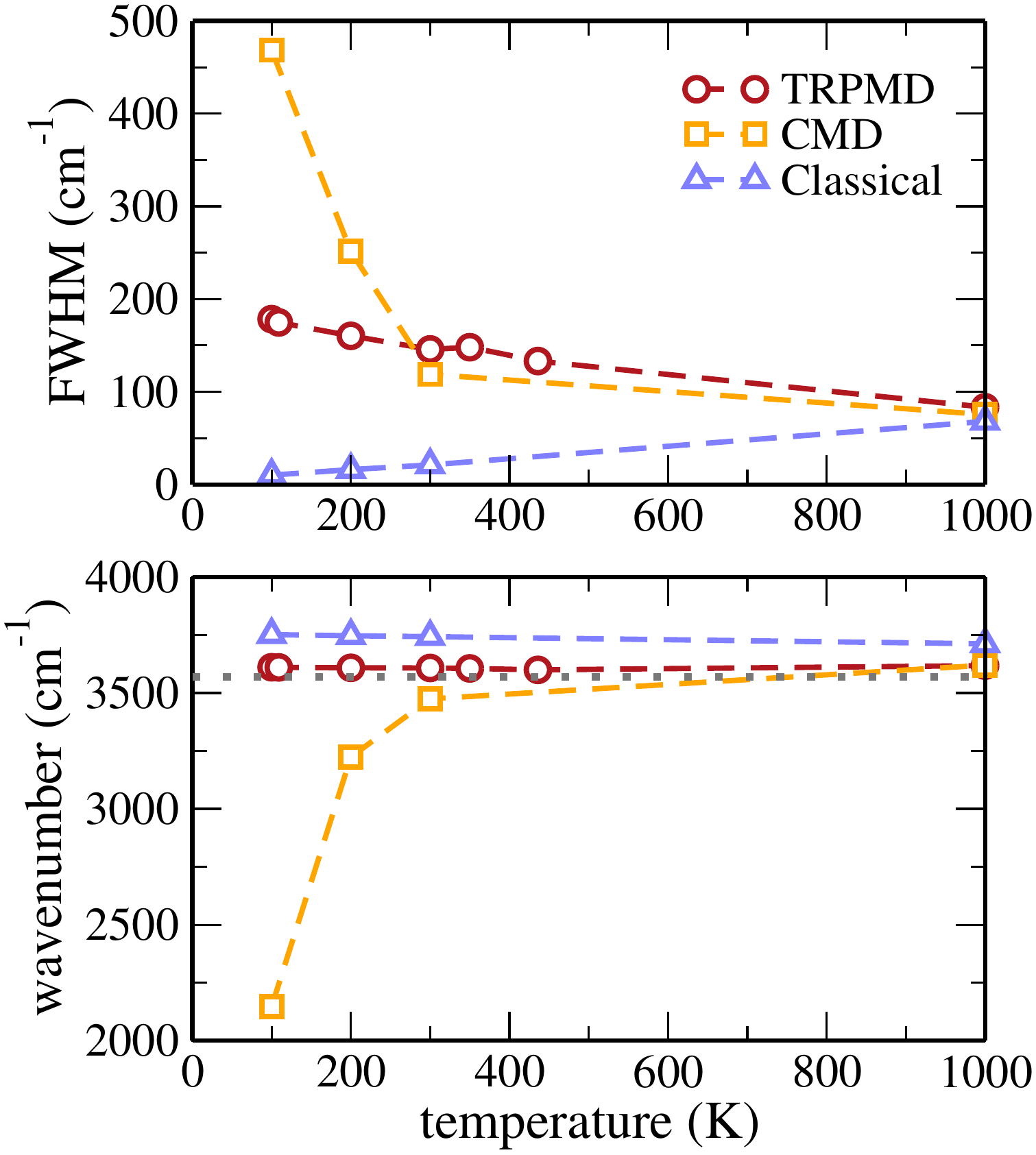}
\caption{Full width at half maximum FWHM (top) and peak position (bottom) as a function of temperature for the CMD and TRPMD spectra shown in Fig. \ref{fig:ohreal-rpmd-cmd-pile}, plus additional data from a classical simulation and points at 1000 K for all three methods. The dotted line in the bottom panel indicates the correct position of the peak at 3568 cm$^{-1}$.}
\end{figure}

The optimally-damped TRPMD results shown in panels (b) and (d) of Fig.~4 are again unaffected by the curvature and resonance problems. The TRPMD stretching peak is within 35 cm$^{-1}$ of the correct fundamental transition frequency, whereas the harmonic vibrational frequency of the molecule (which would be obtained from a classical simulation at low temperature) is blue shifted from the fundamental frequency by 170 cm$^{-1}$. The position of the TRPMD peak is also independent of the simulation temperature, although it does broaden as the temperature is lowered, and more so than in the case of the harmonic OH molecule considered in Fig.~1.

In order to quantify the extent of this broadening, we have fitted Lorentzian curves to the CMD and TRPMD peaks in Fig.~4. The resulting peak positions and full widths at half maximum (FWHM) are shown as a function of temperature in Fig.~5, along with the results of classical simulations and additional calculations at 1000 K. The peaks proved to be almost perfectly Lorentzian in all cases except at 1000 K, where they were slightly asymmetric. From the top panel of Fig.~5 it is clear that, while the TRPMD vibrational peak broadens with decreasing temperature, the broadening is less severe than in the case of CMD. The peak position, shown in the bottom panel, is also stable in the TRPMD and classical simulations, but not in CMD. The classical simulation produces a peak with a FWHM of only 11cm$^{-1}$ at 100 K, but the width increases with temperature, and becomes comparable to the widths of the CMD and TRPMD peaks at 1000 K.

\subsection{The Zundel cation}

\begin{figure*}[htbp]
\centering
\subfigure{
\includegraphics[width=0.435\textwidth]{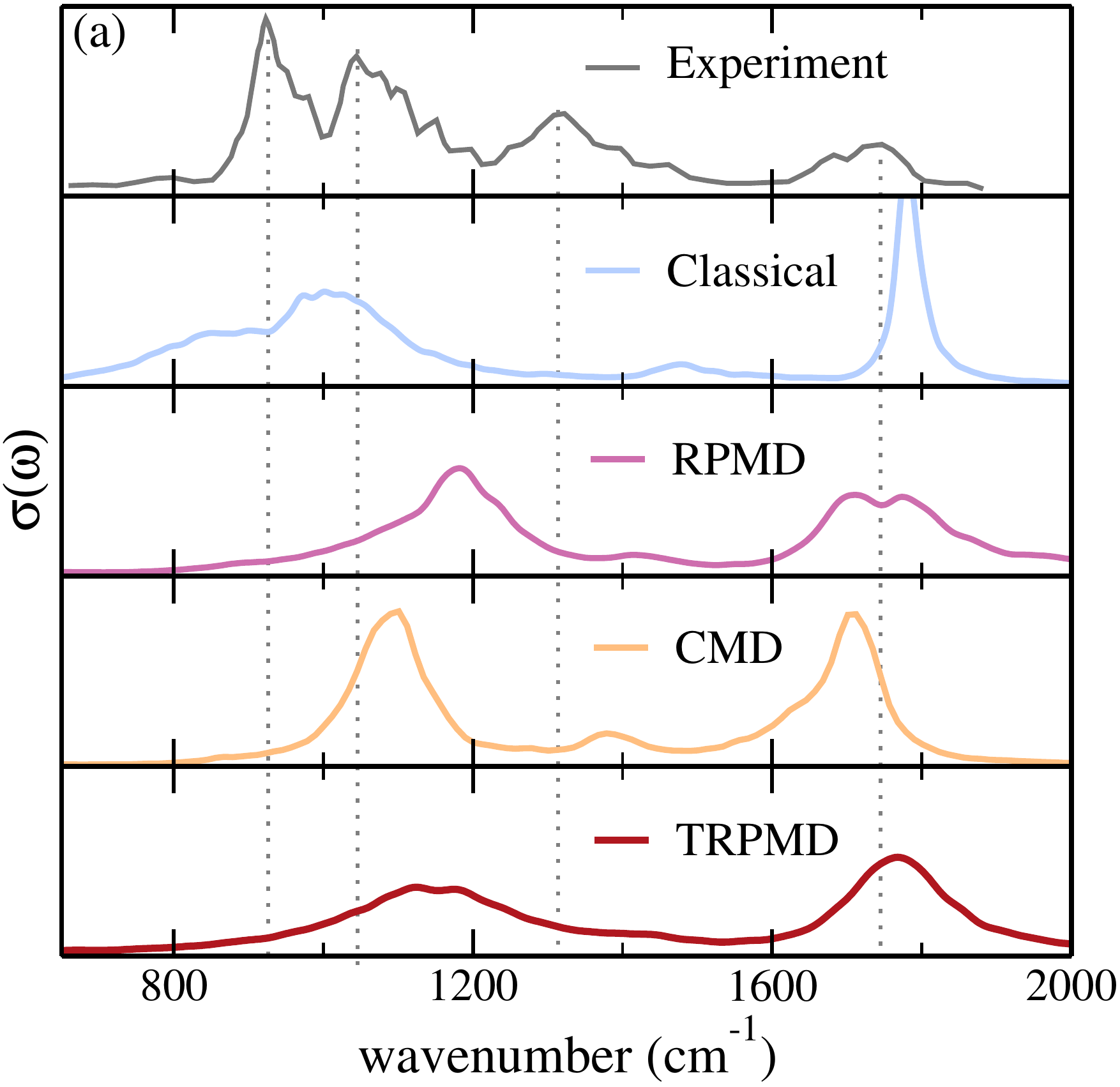}
}
\subfigure{
\includegraphics[width=0.25\textwidth]{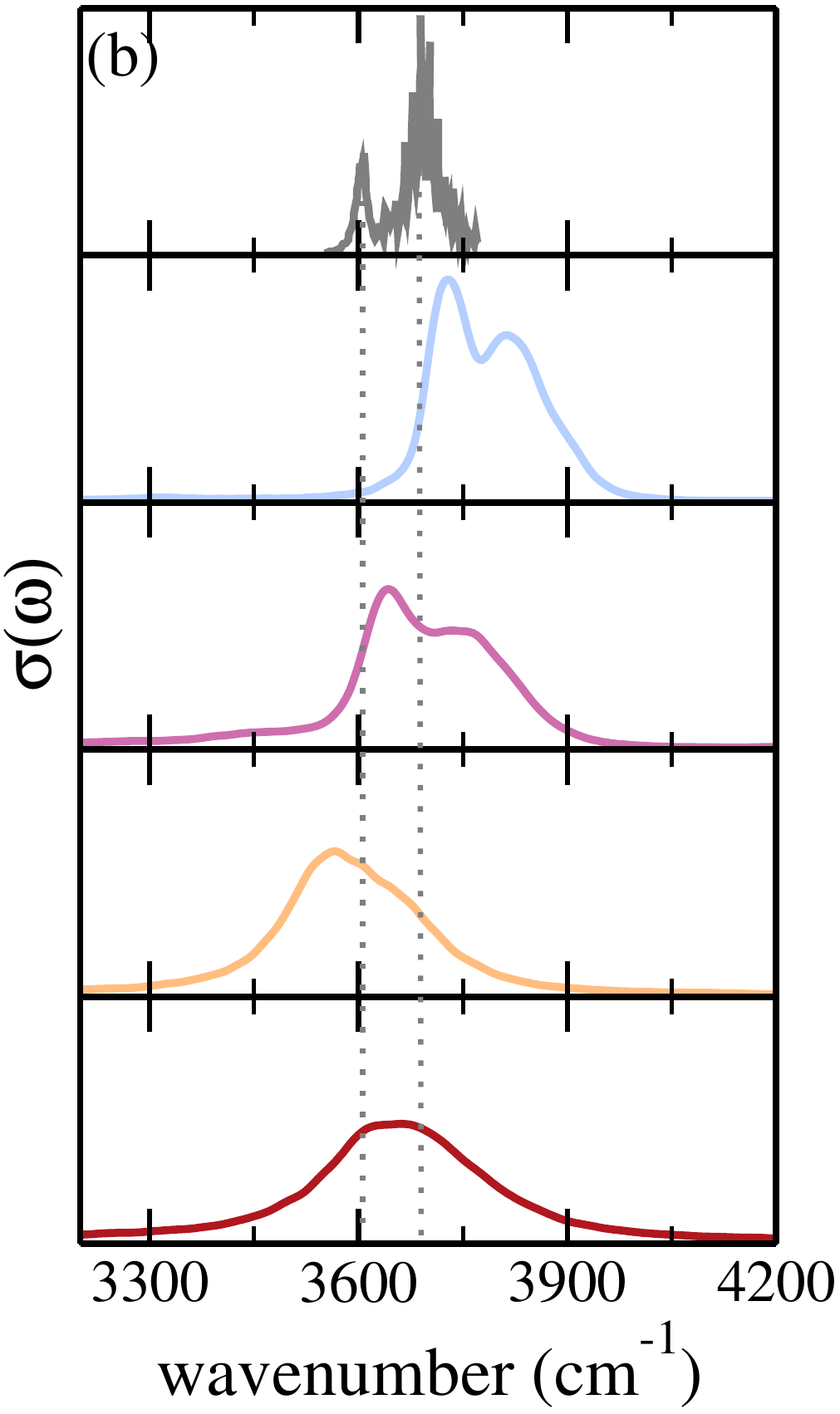}
}
\caption{
(a) Top panel: experimental infrared multi-photon dissociation (IRMPD) spectrum of H$_5$O$_2^+$ in the 800-2000 cm$^{-1}$  region at 100 K from Ref. \cite{asmis03}. The other panels show simulated spectra at this temperature obtained from the Fourier transform of the dipole autocorrelation function using the CCSD(T) parameterised potential energy and dipole moment surfaces of Ref. \cite{huang05}. From top to bottom: classical molecular dynamics, RPMD, CMD, and optimally-damped TRPMD. (b) Top panel: experimental IRMPD spectrum of H$_5$O$_2^+$ in the OH stretch region at 300 K from Ref. \cite{yeh89}. The other panels are the same as in (a), but at 300 K.
}
\end{figure*}

Next, to illustrate the limitations of using any method like RPMD, CMD, or TRPMD to simulate gas phase vibrational spectra, we have considered a significantly more complex test case: the Zundel cation H$_5$O$_2^+$ described by the CCSD(T) potential energy and dipole moment surfaces of Huang, Braams and Bowman.\cite{huang05} The infrared spectroscopy of this cation has been studied extensively in the past both theoretically\cite{vendrell07,neidner-schatteburg08,huang08,kaledin09,baer10,agostini11} and experimentally,\cite{yeh89,asmis03,fridgen04,hammer05,guasco11} and it contains features that one should not even hope to capture using a method that neglects real time quantum interference effects. Among other things, the low temperature spectrum in the shared proton region contains a doublet feature that has been interpreted as a Fermi resonance involving a fourth-order coupling  between the proton transfer mode, the O-O stretching mode, and the H-O-H wagging mode.\cite{vendrell07, neidner-schatteburg08} This feature can only be captured by high levels of theory that include anharmonic effects and a fully quantum mechanical treatment of the nuclear dynamics, such as the multi-configurational time-dependent Hartree (MCTDH) method at 0 K.\cite{vendrell07}

With this caveat in mind, let us first consider the frequency range corresponding to the OH stretch (Fig.~6(b)). The present results were obtained from 200 6 ps trajectories at 300 K, using 16 beads and without removing rotations. Experiments at this temperature~\cite{yeh89} show two distinct peaks at about 3600 and 3700 cm$^{-1}$. As in the case of the anharmonic OH discussed above, the classical spectrum is blue shifted from these peaks by almost 150~cm$^{-1}$. The RPMD spectrum is closer to the experiment, but it has an incorrect profile and a broad tail towards lower frequencies. In a multi-dimensional problem such as this, there can be resonances with the internal modes of the ring polymer associated with a variety of different molecular vibrational modes, which makes it hard to assess whether and how the RPMD spectrum is contaminated by resonances. The CMD spectrum is red shifted with respect to the experiment, and consists of a single broad peak. Given the high accuracy of the potential energy surface, we believe that the red shift can be attributed to the curvature problem, and we have checked this by decreasing the temperature: the red shift then becomes even more pronounced in the same way as in Figs.~1 and~4. Finally, the optimally-damped TRPMD spectrum shows neither resonances nor any curvature-related red shift. The position of the absorption band is in reasonable agreement with the experiment, although the significant broadening that was also observed for the anharmonic OH molecule washes out the doublet structure and yields a single peak centred at about 3650 cm$^{-1}$.

We have also run simulations at 100 K using 64 beads to compare RPMD, CMD and TRPMD with an experiment performed in the intermediate frequency region of the spectrum~\cite{asmis03} (Fig.~6(a)). All three methods generate very broad peaks in this region, and the RPMD spectrum displays oscillations in its bending peak at around 1750 cm$^{-1}$ that can probably be attributed to a resonance effect. Interestingly, the CMD bending peak is red-shifted relative to experiment, suggesting that at such low temperatures the curvature problem can also affect lower frequency modes than OH stretching vibrations. The bending peak from the optimally-damped TRPMD calculation is closest to the experiment and shows no signs of resonances. As discussed above, none of the methods investigated here can reproduce the experimental doublet at around 1000 cm$^{-1}$, which is due to a subtle Fermi resonance effect.\cite{vendrell07,neidner-schatteburg08} The TRPMD calculation simply produces a single broad absorption band that spans the range of all three experimental peaks between 800 and 1400 cm$^{-1}$, and the RPMD and CMD calculations do not fare any better. 

\subsection{Liquid water}

Having highlighted a situation in which none of the present methods is really satisfactory, it is only fair to end with an application for which methods like RPMD and CMD are designed: the simulation of liquid water. This is a fundamentally different problem because the exact quantum mechanical dipole autocorrelation function of a liquid decays to zero at long times, resulting in a continuous spectrum. The classical approximation to this spectrum (to which both RPMD and CMD are pinned at high temperatures) becomes exact in the limit where $\beta\hbar\omega\ll 1$, which is not the case for the discrete spectrum of a small gas phase molecule. Methods like RPMD and CMD (and indeed even classical MD) are therefore on far firmer ground for the simulation of liquids than they are for isolated molecules.

With this in mind, we have used classical MD, RPMD, CMD and optimally-damped TRPMD to calculate the dipole absorption spectrum of the q-TIP4P/F water model.\cite{habershon09} This water model was chosen because it reproduces a wide variety of structural, thermodynamic and dynamical properties of liquid water in path integral simulations, including the vibrational band frequencies which were optimised in the parameterisation of the model using a partially-adiabatic\cite{hone06} approximation to CMD. Because it is a simple point charge model rather than a polarisable model, it is missing the induced dipole contribution to the change in dipole moment along the OH stretch, which is known to make a significant contribution to the intensity of the OH stretching band. However, the frequency of the stretching band in the model is roughly correct, and the same is true of the intramolecular bending and intermolecular librational bands.\cite{habershon09}

In this case, since the volume $V$ of the system is well defined, it is possible to calculate the absolute signal\cite{habershon08}
$$
n(\omega)\alpha(\omega) = {\pi\beta\omega^2\over 3cV\epsilon_0}\tilde{I}(\omega), \eqno(33)
$$
where $n(\omega)$ is the frequency-dependent refractive index of the liquid, $\alpha(\omega)$ is its Beer-Lambert absorption coefficient, and $\tilde{I}(\omega)$ is the Fourier transform of its Kubo-transformed dipole autocorrelation function. The present calculations were performed with the i-PI\cite{ceriotti14} path integral code, using LAMMPS\cite{plimpton95} as the backend to calculate energies and forces. We used 216 water molecules in a periodically replicated simulation box with the experimental density at 300 K. The classical results were averaged over 8 independent 100 ps trajectories initiated from an equilibrated simulation, with a time step of 0.25 fs. The optimally-damped TRPMD calculations were performed in exactly the same way, but with $n=32$ ring polymer beads. In the CMD calculations, the masses of the internal modes of the free ring polymer were adjusted to bring them to a frequency of 16000 cm$^{-1}$, and we decreased the integration time step to 0.025 fs to correctly integrate the rapid oscillations of these modes. RPMD calculations are typically done by averaging over hundreds of short trajectories,\cite{craig04} with the momenta resampled from the Boltzmann distribution between each one in order to overcome the non-ergodicity of the microcanonical ring polymer dynamics.\cite{hall84} However, since RPMD is the $\lambda\to 0$ limit of TRPMD, it is also possible to run RPMD in the same way as TRPMD with an under-damped thermostat ($\lambda=0.001$) that is weak enough not to disrupt the dynamics but strong enough to enhance the ergodicity. This is how we performed the present RPMD calculations, again using 8 independent 100 ps trajectories with a time step of 0.25 fs.

\begin{figure*}[htbp]
\centering
\subfigure{
\includegraphics[width=0.5\textwidth]{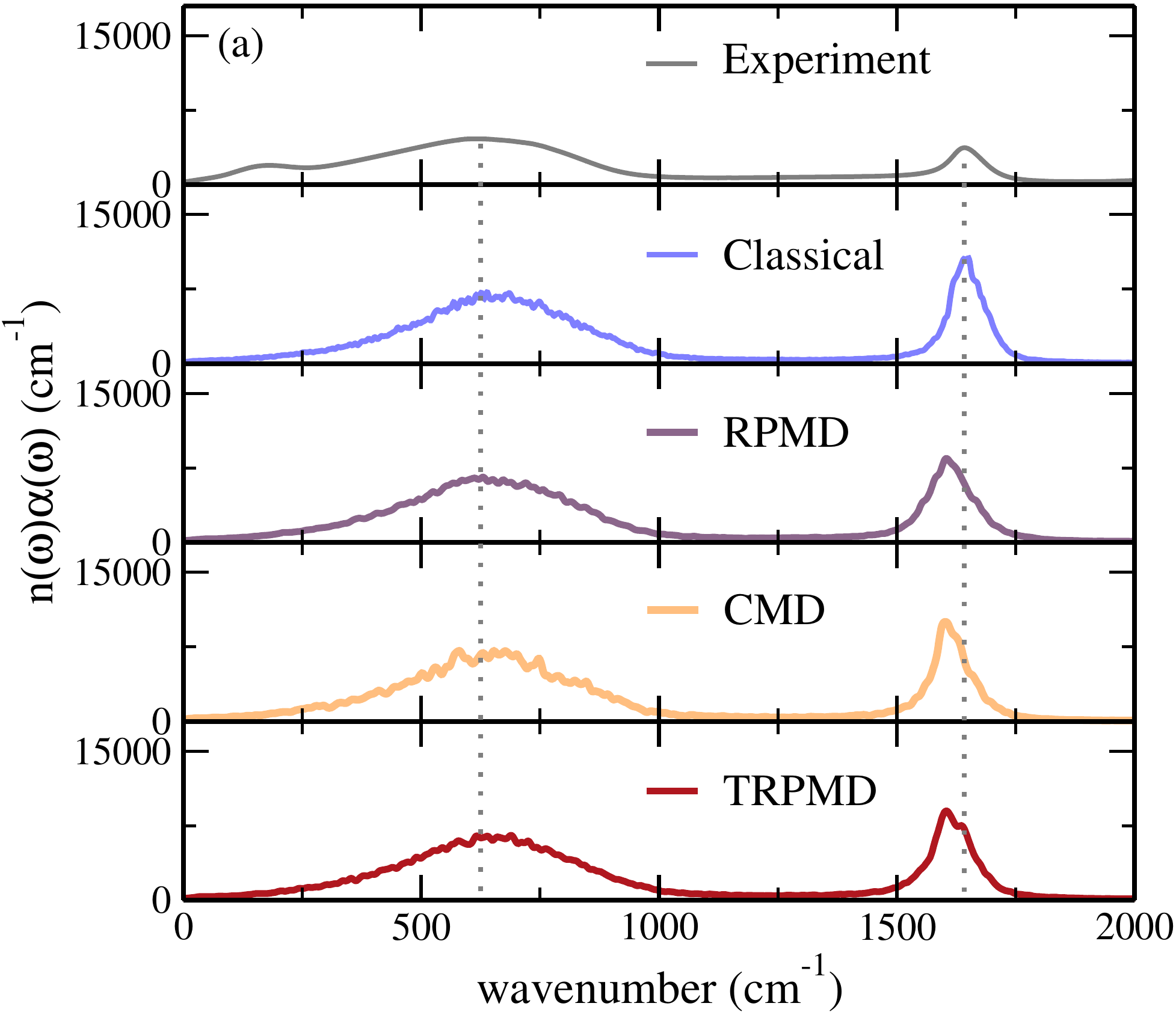}
}
\subfigure{
\includegraphics[width=0.23\textwidth]{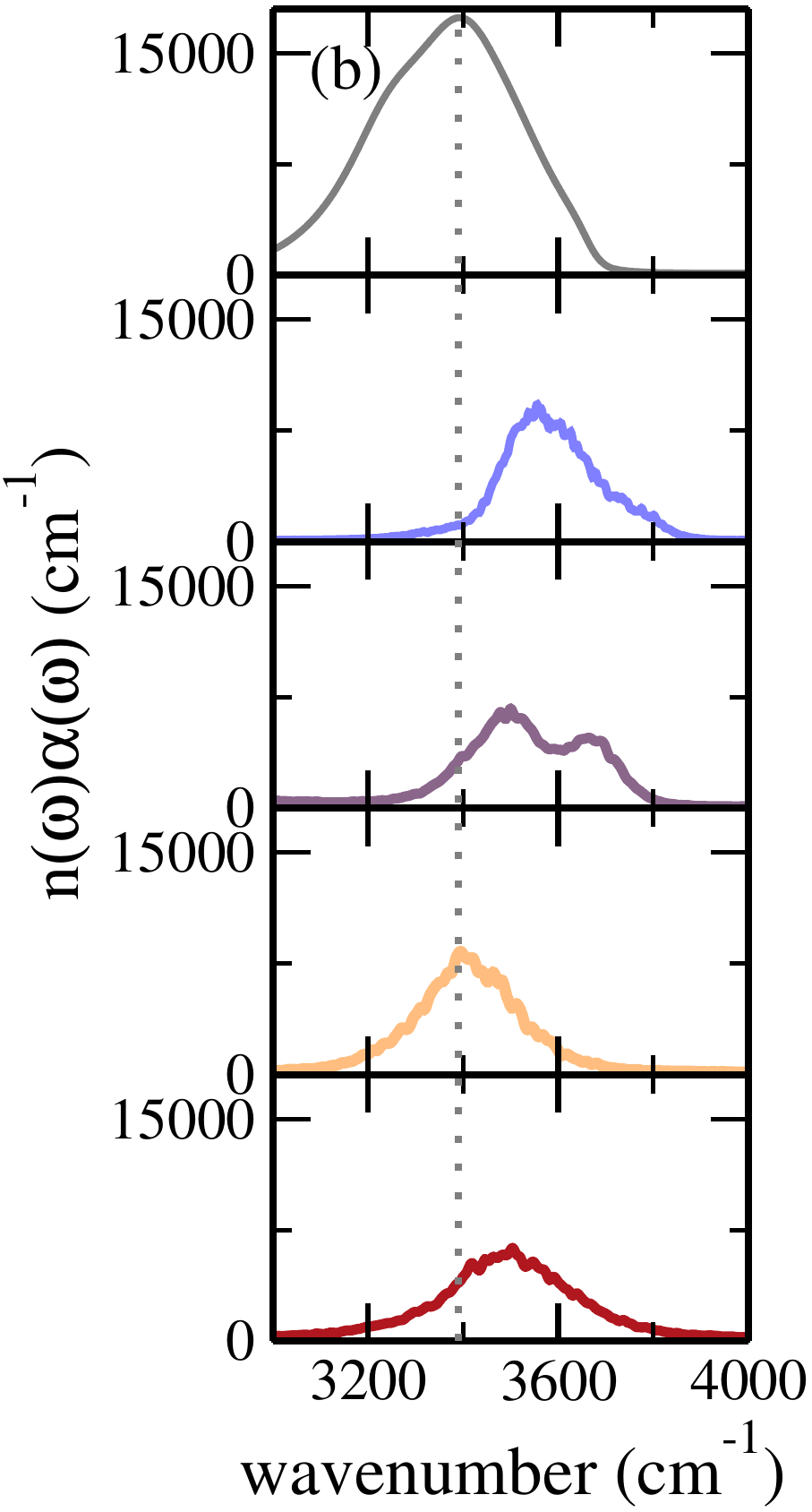}
}
\caption{
(a) Top panel: experimental infrared absorption spectrum of room temperature liquid water from Ref.~\cite{bertie96}
between 0 and 2000 cm$^{-1}$. The other panels show simulated spectra at this temperature obtained from the Fourier transform of the dipole autocorrelation function using the q-TIP4P/F water model.\cite{habershon09} From top to bottom: classical molecular dynamics, RPMD, CMD, and optimally-damped TRPMD. (b) Same as in (a), but in the region between 3000 and 4000 cm$^{-1}$.}
\end{figure*}

The results of these calculations are compared with the experimental absorption spectrum of liquid water\cite{bertie96} in Fig.~7. All four simulations give very similar spectra in the low frequency librational region, where they are in good agreement with the experiment. In the intermediate frequency bending region, the peaks of the three path integral methods are again seen to be in agreement, but they are red shifted from the classical bending peak by around 35 cm$^{-1}$. Given the good agreement between the three path integral methods, the absence of any evident curvature problem in the CMD bending peak, and the absence of any apparent resonance problem in the RPMD bending peak, we conclude that this is the correct anharmonic red shift of the bending band for the q-TIP4P/F water model.
 
The differences between the path integral methods occur in the high frequency OH stretching region. Here the RPMD calculation shows clear indications of the resonance problem, although the resonances are not as pronounced as those seen previously using the TTM3-F\cite{fanourgakis08} water model,\cite{habershon08} which has a non-linear dipole moment operator that accentuates the resonance coupling. The resonances are washed out in the TRPMD spectrum, which exhibits a single peak that is red shifted from the classical peak by some 75 cm$^{-1}$. The CMD peak is red shifted by a further 80 cm$^{-1}$. This peak is clearly in the best agreement with the experiment, but this is simply because the q-TIP4P/F potential was parameterised to agree with this experiment in a (partially adiabatic) CMD calculation.\cite{habershon09}

The issue of the curvature problem in CMD simulations of liquid water has been the subject of some debate. Paesani and Voth\cite{paesani10} have argued, on the basis of partially adiabatic CMD simulations with a weak adiabatic decoupling parameter of a solvated HDO molecule in H$_2$O and D$_2$O, and comparisons of the OD and OH stretching peaks of the solute with those given by mixed quantum-classical calculations, that the curvature problem is unlikely to play a significant role in simulations of liquids like q-TIP4P/F with anharmonic stretching potentials. On the other hand, Ivanov {\em et al.}\cite{ivanov10} have argued that the curvature problem is indeed present in the stretching bands of CMD simulations of neat liquid HDO, H$_2$O, and D$_2$O with both harmonic and anharmonic OH stretching potentials. 

In the present simulations, a comparison with the position of the TRPMD OH stretching peak, which we would not expect on the basis of the other vibrational problems we have considered above to be shifted very far from the correct position, suggests that the CMD peak does show some evidence of the curvature problem. This is not to say that the TPRMD OH stretching frequency is exactly right, and indeed it may well not be: recall the $\sim 35$ cm$^{-1}$ blue shift that this method gives from the correct fundamental transition frequency of the anharmonic OH molecule (Fig.~4). In view of this uncertainty, which prevents us from drawing any more definite conclusions here, we believe it would be worthwhile in a future study to compare the present CMD and TRPMD results for q-TIP4P/F liquid water with those of a more accurate (and expensive) quantum mechanical calculation, for example using the recently-developed local monomer approximation.\cite{wang11,liu12}

Finally, Table~\ref{tab:coefficients} compares the diffusion coefficients and various rotational correlation times $\tau^{\eta}_l$ obtained from the present classical, RPMD, CMD, and optimally-damped TRPMD simulations. One sees that all three path integral methods are in qualitative agreement in predicting a 12\% increase in the diffusion coefficient and a $\sim 20$\% decrease in the rotational correlation times relative to the classical simulation, which is consistent with the earlier results for the q-TIP4P/F water model reported in Ref.~\cite{habershon09}. Indeed the differences between the path integral methods for these zero-frequency dynamical properties are mostly within the statistical error bars of our simulations (obtained from 8 independent 100 ps trajectories for each method).

\setlength{\extrarowheight}{5pt}
\begin{table}[ht!]
\caption{Dynamical properties of liquid water at 300 K obtained from classical, RPMD, CMD and TRPMD  simulations of the q-TIP4P/F water model. $D$ is the diffusion coefficient and $\tau_l^{\eta}$ is the $l$th order orientational relaxation time for molecular axis $\eta$.}
\label{tab:coefficients}
\begin{tabular*}{0.48\textwidth}{@{\extracolsep{\fill} } ccccc}
\hline
\hline
    & Classical & RPMD & CMD & TRPMD \\
\hline
$D_{\text{H}_2\text{O}}$ (\AA$^2$ ps$^{-1}$) & 0.194(5) & 0.218(3) & 0.219(2) & 0.217(2) \\
$\tau^{\text{HH}}_1$ (ps)                                 & 6.1(2)     &  5.0(1)    & 5.1(1)    &  5.1(1)    \\
$\tau^{\text{OH}}_1$ (ps)                                 & 5.8(2)     &  4.8(1)   &  4.9(1)    & 4.8(1)     \\
$\tau^{\mu}_1$ (ps)                                          & 5.2(2)     &  4.4(1)   &  4.5(2)     & 4.4(1)       \\
$\tau^{\text{HH}}_2$ (ps)                                 &  2.55(8)  &  2.15(7)   & 2.17(5) & 2.08(6)    \\
$\tau^{\text{OH}}_2$ (ps)                                 &  2.23(7)   &  1.85(5)   & 1.90(4) & 1.81(4)  \\
$\tau^{\mu}_2$ (ps)                                          & 1.87(8)   &  1.42(4)   & 1.49(4) & 1.45(5)  \\
\hline
\hline
\end{tabular*}
\end{table}

\section{Concluding remarks}

In this paper, we have shown that it is possible to apply an internal mode thermostat to RPMD without altering any of its established properties (see Sec.~II), and that the resulting method -- which is halfway between CMD and RPMD -- is significantly better than either of these alternatives for vibrational spectroscopy (see Sec.~III). The results are still not perfect, and we would certainly not recommend using any of the methods we have considered here to simulate the spectroscopy of small gas phase molecules. However, for more complex systems, ranging from large biomolecules to liquids, optimally-damped TRPMD might at least be expected to provide a reasonable prediction of the anharmonic shifts in vibrational fundamental bands, at a modest computational cost.

Since TRPMD is no less valid (or {\em ad hoc}!) than standard RPMD, and we have shown here that it works better for vibrational spectroscopy, it is natural to ask whether it might also offer an improvement for other applications, such as the calculation of diffusion coefficients and chemical reaction rates.  Does the TRPMD velocity autocorrelation function of a quantum liquid such as para-hydrogen satisfy imaginary time moment constraints more accurately than the RPMD velocity autocorrelation function? Their transition state theory limits are the same, but are the TRPMD transmission coefficients of some simple chemical reactions (for which the exact rates can be computed for comparison) more accurate than the RPMD transmission coefficients? These are interesting questions for future work.

Finally, we should stress that the PILE thermostat we have considered here is by no means unique. By moving to a coloured noise (generalised Langevin equation) thermostat, in the form of a Langevin thermostat in an extended momentum space, it may well be possible to find something better. There is clearly a tremendous amount of leeway in the construction of extended phase space approximations to quantum dynamics, and it is almost certain that we have yet to find the best one. Methods like CMD and RPMD have simply scratched the surface of what might be possible.

\begin{acknowledgements}
We are grateful to David Wilkins for providing us with a code for calculating rotational correlation times and to Tom Markland for extensive comments on a preliminary draft of this manuscript. This work was supported by the Deutsche Forschungsgemeinschaft (DFG), the Wolfson Foundation, and the Royal Society.  
\end{acknowledgements}

\end{document}